\documentclass[a4paper,conference]{IEEEtran}
\usepackage[numbers,sort&compress]{natbib}

\ifCLASSINFOpdf
   \usepackage[pdftex]{graphicx}
   \graphicspath{{../pdf/}{../jpeg/}}
   \DeclareGraphicsExtensions{.pdf,.jpeg,.png}
\else
   \usepackage[dvips]{graphicx}
   \graphicspath{{../eps/}}
   \DeclareGraphicsExtensions{.eps}
\fi
\usepackage{amsmath}
\usepackage{algorithmic}

\usepackage{array}

\ifCLASSOPTIONcompsoc
  \usepackage[caption=false,font=normalsize,labelfont=sf,textfont=sf]{subfig}
\else
  \usepackage[caption=false,font=footnotesize]{subfig}
\fi

\usepackage{stfloats}
\usepackage{url}

\usepackage{amsfonts}
\usepackage{algorithm}
\usepackage{xcolor}

\newsavebox{\ieeealgbox}

\begin{document}

\title{Assortative-Constrained Stochastic Block Models}

\author{\IEEEauthorblockN{Daniel Gribel}
\IEEEauthorblockA{Departamento de Inform\'atica\\
Pontif\'icia Universidade Cat\'olica\\do Rio de Janeiro\\
Rio de Janeiro, Brazil, 22451-900\\
Email: dgribel@inf.puc-rio.br}
\and
\IEEEauthorblockN{Thibaut Vidal}
\IEEEauthorblockA{Departamento de Inform\'atica\\
Pontif\'icia Universidade Cat\'olica\\do Rio de Janeiro\\
Rio de Janeiro, Brazil, 22451-900\\
Email: vidalt@inf.puc-rio.br}
\and
\IEEEauthorblockN{Michel Gendreau}
\IEEEauthorblockA{Department of Mathematics and\\Industrial Engineering\\
Polytechnique Montr\'eal\\
Montr\'eal, Canada, P.O. Box 6079\\
Email: michel.gendreau@cirrelt.net}}


%


\maketitle

\begin{abstract}
Stochastic block models (SBMs) are often used to find \emph{assortative} community structures in networks, such that the probability of connections within communities is higher than in between communities. However, classic SBMs are not limited to assortative structures. In this study, we discuss the implications of this model-inherent indifference towards assortativity or disassortativity, and show that this characteristic can lead to undesirable outcomes for networks which are presupposedy assortative but which contain a reduced amount of information. To circumvent this issue, we introduce a constrained SBM that imposes strong assortativity constraints, along with efficient algorithmic approaches to solve it. These constraints significantly boost community recovery capabilities in regimes that are close to the information-theoretic threshold. They also permit to identify structurally-different communities in networks representing cerebral-cortex activity regions.
\end{abstract}

%
\IEEEpeerreviewmaketitle

\section{Introduction}

Community detection methods hold a central place in machine learning, with an extensive range of applications related to sociological behavior, protein interactions, image segmentation, and gene expressions analysis \cite{Abbe2017}. In most of these applications, the actual classes of the nodes in the network are unknown, but pairwise relations between nodes are exploited to identify communities.

Fitting the parameters of a stochastic block model (SBM) \cite{Holland1983, Nowicki2001} to a given graph is a prominent way of searching for communities. The canonical SBM assumes that each node belongs to one block (representing a community) and that the expected number of edges between two nodes depends only on the blocks to which they belong. Thus, the model only assumes that nodes within each block are statistically equivalent in their connectivity patterns. Several variations of the standard SBM were also introduced to overcome some of its limitations. The degree-corrected SBM (DC-SBM) introduced by \citet{Karrer2011}, in particular, allows non-uniform node degree distributions, making block modeling more representative of real-world networks.

\begin{figure}[htbp]
    \centering
    \vspace*{-0.3cm}
    \captionsetup{justification=centering}
    \subfloat[Optimal solution $\log \mathcal{L} = -2.7616$]{{\includegraphics[width=2.72cm]{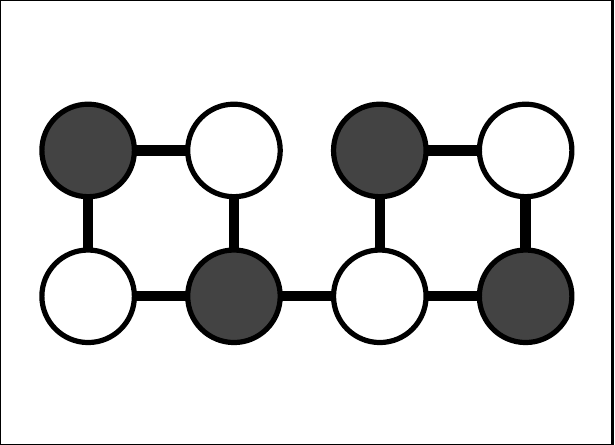}}}%
   \hfill
    \subfloat[2$^\text{nd}$ best solution $\log \mathcal{L} =-5.3193$]{{\includegraphics[width=2.72cm]{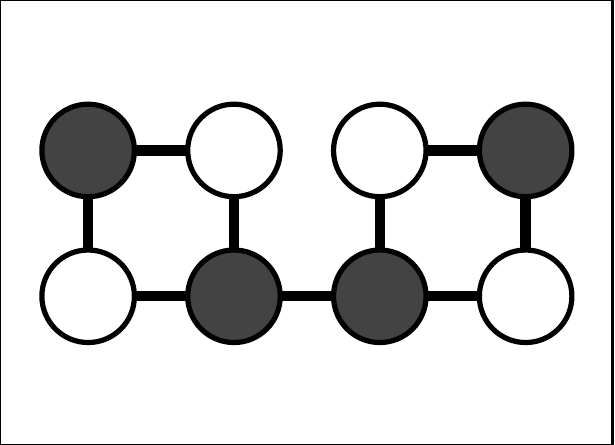}}}%
    \hfill
    \subfloat[3$^\text{rd}$ best solution $\log \mathcal{L} =-5.9012$]{{\includegraphics[width=2.72cm]{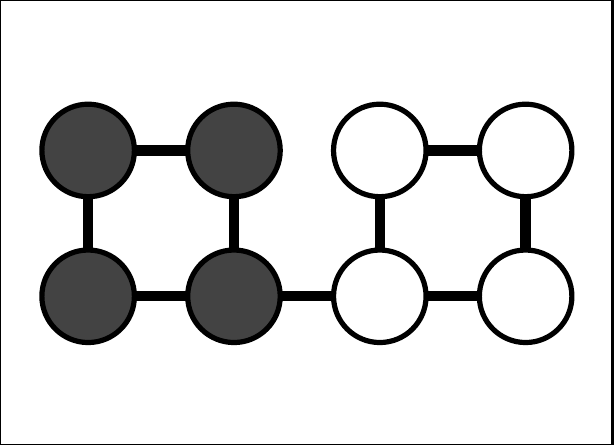}}}%
    \caption{The three best solutions in a small example case with eight nodes and two communities. The two best solutions in terms of maximum likelihood are disassortative, whereas the third (c) is assortative.}
    \label{fig:AssortativeExample}%
\end{figure}

Broadly speaking, a solution for community detection (represented as a partition of the node set into communities) is \emph{assortative} when connections within communities are more frequent than in between communities, it is \emph{disassortative} when connections within communities are less frequent than in between communities, and finally it is \emph{non-assortative} if no such relation exists among all communities. 
SBM-based community detection approaches are agnostic to the assortativity of their solutions. They allow to search for solutions with a pre-defined number of communities, and can indifferently model assortative and disassortative structures. This modeling capability can be viewed as an asset but also as a weakness. Indeed, SBMs are often used in contexts in which users expect assortative solutions. In the most dramatic situations, non-assortative solutions might go under the radar and lead to mistakes of interpretation. In other cases, non-assortative solutions with a better likelihood may substitute the assortative solutions which were originally sought (Figure~\ref{fig:AssortativeExample}). This later situation is especially prevalent in case studies involving sparse graphs, or with lightly assortative structures which challenge detection algorithms.

In this work, we propose a variant of the DC-SBM which includes user knowledge about assortativity. We incorporate this information by setting assortativity constraints on the DC-SBM parameter set. Indeed, if the user expects an assortative solution due to the characteristics of the application case, it is plausible to guide the convergence of the model via additional constraints. We show that the resulting constrained likelihood-maximization model can be solved efficiently with an iterative method based on local-search and interior-point algorithms. Our computational experiments show that the assortativity constraints prevents the search from converging towards spurious non-assortative local minima, especially for sparse networks, and that they contribute to identify different solution structures in application cases related to the analysis of the brain cortex. The key contributions of this work are therefore the following.
\begin{enumerate}
\item[1)] We introduce a DC-SBM variant which incorporates assortativity constraints to represent prior user knowledge;
\item[2)] We propose an efficient solution approach based on local optimization and interior-point algorithms for this model;
\item[3)] Through extensive computational experiments, we discuss the practical implications of this constrained model and identify the regimes in which it contributes to improve community detection practice.
\end{enumerate}

\section{Related Works}

SBMs are commonly used to extract meaningful information from complex networks. The classical SBM is also a natural modeling choice for community detection \cite{Abbe2017} and a generalization of modularity maximization \cite{Newman2016}. The surveys of \citet{Abbe2017} and \citet{Lee2019} discuss key results regarding recovery requirements and solution algorithms. 
Different types of algorithms can be used to fit SBMs, based on Markov Chain Monte Carlo (MCMC) approaches \cite{Mcdaid2013,Nowicki2001,Peixoto2019}, variational inference \cite{Wang2017,Airoldi2008}, belief propagation \cite{Decelle2011}, spectral clustering \cite{Lei2015,Qin2013,Rohe2011}, and semidefinite programming \cite{Cai2015,Chen2012}, among others.

To date, few works have considered the possibility of incorporating prior information on assortativity. \citet{Moore2011} studied a SBM in which the edge probabilities within and in between communities follow a Beta prior. The hyperparameters defining the Beta distributions drive the degree of assortativity in the graph. Yet, according to their experiments, these priors dominate only in small or sparse data sets, otherwise they tend to 
wash out.

The Assortative Mixed Membership SBM (a-MMSB) introduced by \citet{Gopalan2012} considers soft node-to-community assignments and includes a latent parameter describing community strength, representing how tightly nodes are connected within each group. Edges are assumed to be drawn from a Bernoulli distribution centered around the community strength if the nodes belong to the same group. Otherwise, the distribution is centered around a small value. A variational inference approach is used to fit the model. \citet{Li2016} pursued this research line by proposing a scalable MCMC method using a stochastic gradient algorithm for posterior inference in the a-MMSB. 

\citet{Lu2019} finally proposed a regularized variant of the DC-SBM, using a prior to regularize the observed in-degree ratio of each node. In practice, this adaptation turns out to penalize high-degree nodes with many connections to other communities. The new parameter is adjusted to control the assortativity level, and a MCMC algorithm is used to infer the block assignments.

The aforementioned models aim to better fit assortative networks, but they are either dependent on ad-hoc parameters which are difficult to scale \citep{Gopalan2012,Lu2019}, or of limited effect for larger graphs \citep{Moore2011}. In light of these works, we decided to explore a different approach, which consists in guiding the search towards assortative structures via constraints in the SBM parameters. To fit our model, we propose effective algorithms for the resulting constrained maximum-likelihood optimization problem.

\section{Background and Notations}

\subsection{Degree-Corrected Stochastic Block Model (DC-SBM)}

In its most fundamental form, the DC-SBM considers $N$ nodes allocated to $K$ groups. We assume that the number of edges between a pair of nodes $(i,j)$ depends only on the groups to which the nodes belong and on their degrees \cite{Newman2016}. Finding the latent membership of nodes corresponds to finding the block-model parameters that best fit the observed graph \cite{Abbe2017}. For an observed adjacency matrix $A \in \mathbb{N}^{N, N}$ representing a graph with $m$ (possibly weighted) edges, the log-likelihood function of the DC-SBM is calculated as \cite{Karrer2011}:
\begin{equation}
    \log P(A | \Omega,\!Z) = \frac{1}{2} \sum_{rs}^{K}\!\sum_{ij}^{N} \left ( A_{ij} \log(\omega_{rs}) - \frac{k_i k_j}{2m} \omega_{rs}  \right ) \!z_{ir} z_{js}.
\label{DCSBM}
\end{equation}

In this equation, $k_i$ is the degree of node $i$, and $m$ is the total number of edges. Variables $Z \in \{0, 1\}^{N, K}$ represent the binary community assignments, in such a way that $z_{ir} = 1$ indicates that node $i$ is assigned to group $r$. $\Omega$ is a symmetric $K \times K$ edge probability matrix. Each element $\omega_{rs}$ of $\Omega$ corresponds to the expected number of edges between any two points in groups $r$ and $s$. The expected number of edges between nodes $i$ and $j$ is $\frac{k_i k_j}{2m} \omega_{rs}$, for $z_{ir} = 1$ and $z_{js} = 1$. If we fix the assignment $Z$, then the (unconstrained) maximum-likelihood for each parameter $\omega_{rs}$ can be estimated by differentiation:
\begin{equation}
\label{OmegaMax}
	\hat{\omega}_{rs} = \frac{2m \cdot m_{rs}}{\kappa_r \kappa_s},
\end{equation}
where $\smash{m_{rs} = \sum_{ij}^{N} A_{ij} z_{ir} z_{js}}$ is the number of edges between groups $r$ and $s$, and $\kappa_{r} = \sum_{i}^{N} k_i z_{ir}$ is the sum of the degrees of nodes in group $r$. If we substitute $\hat{\omega}_{rs}$ in Equation \eqref{DCSBM}, we obtain the following log-likelihood function:
\begin{equation}
\log P(A | Z) = \frac{1}{2} \sum_{rs}^{K} \sum_{ij}^{N}\left ( A_{ij} \log \left ( \frac{m_{rs}}{\kappa_r \kappa_s} \right )  \right ) z_{ir} z_{js},
\end{equation}
in which we dropped the terms that do not involve $Z$.

\subsection{Planted Partition Model and Modularity}
\label{sec:presentation-PPM}

The Planted Partition Model (PPM) is a special case of the standard SBM with only two parameters describing the blocks: $\omega_{rs} = \omega_\textsc{in}$ if $r = s$, and $\omega_{rs} = \omega_\textsc{out}$ if $r \neq s$. \citet{Newman2016} shows that maximizing the likelihood of the PPM is equivalent to maximizing modularity. Modularity optimization maximizes the difference between the observed graph and a random graph where edges are reinserted randomly and the degrees of each node is preserved. As a consequence, it results in maximizing the number of edges within groups, leading to assortative solutions. However, modularity maximization is also subject to strong limitations: beyond its inability to define the number $K$ of communities, the model assumes that all communities have similar statistical properties~\cite{Newman2016}. This is a major issue when the distribution of edges between the blocks varies significantly.

\section{Assortative-Constrained SBM}
\label{sec:AC-SBM}

We now introduce the assortative-constrained degree-corrected SBM (AC-DC-SBM) along with an efficient algorithm to fit it by maximum likelihood. Following \citet{Amini2018}, two main notions of assortativity can be distinguished for block models:\\

\noindent
\textbf{Strong assortativity.} All diagonal terms of $\Omega$ are greater or equal than all off-diagonal terms:
	\begin{align}
	\omega_{qq} \geq \omega_{rs} \quad \forall \, q,r,s  \in \{1,\dots,K\}, r \neq s. \label{strong-assortativity-condition}
	\end{align}

\noindent
\textbf{Weak  assortativity.} Each diagonal term of $\Omega$ is greater or equal than the other terms in its row:
	\begin{align}
	\omega_{qq} \geq \omega_{qs} \quad  \forall \, q,s  \in \{1,\dots,K\}.
\label{weak-assortativity-condition}
	\end{align}

Other types of assortativity constraints may be considered with simple adaptations of our algorithm, e.g., imposing a lower bound on the number of blocks satisfying Condition~(\ref{weak-assortativity-condition}). In this study, we will use the strongest definition of assortativity based on Condition~(\ref{strong-assortativity-condition}). With these constraints, the log-likelihood maximization model becomes:
\begin{subequations}
\begin{alignat}{3}
\max_{\Omega, Z, \lambda} & \quad \frac{1}{2} \sum_{rs}^{K} \sum_{ij}^{N} \left (A_{ij} \log(\omega_{rs}) - \frac{k_i k_j}{2m} \omega_{rs}  \right ) z_{ir} z_{js} \label{AC-DC-SBM01}\\
\textrm{s.t.} & \quad \omega_{qq} \geq \lambda \quad \forall q \in \{1,\dots,K\} \label{AC-DC-SBM02}\\
& \quad \omega_{rs} \leq \lambda \quad \forall r,s \in \{1,\dots,K\}, r \neq s \label{AC-DC-SBM03}\\
& \quad \omega_{rs} \geq 0 \quad \forall r,s \in \{1,\dots,K\}, \label{AC-DC-SBM04}
\end{alignat}
\end{subequations}
where $\lambda$ represents a continuous variable acting as a threshold.

It is important to note that the assortativity constraints only apply on the block-model parameters $\Omega$. This does not completely eliminate the possibility of a disassortative partition as represented by $Z$, but strongly penalizes its log-likelihood in comparison to other assortative solutions.

\subsection{Likelihood Maximization}

We introduce an iterative algorithm to solve \mbox{(\ref{AC-DC-SBM01}--\ref{AC-DC-SBM04})}. This algorithm starts with a random initial solution and proceeds by iteratively evaluating each possible relocation of a node to a different community. Each such relocation is only applied if its application combined with an optimal update of~$\Omega$ results into an improvement of the likelihood. As such, the evaluation of each relocation may require the solution of a small constrained convex optimization subproblem with $K^2$ variables and constraints to find an optimal~$\Omega$ for the new partition. For the classical DC-SBM, the optimal~$\Omega$ is simply obtained via Equation~(\ref{OmegaMax}). This is however, no longer true for the AC-DC-SBM due to the assortativity constraints. As described in Algorithm~\ref{LocalSearch-SBM}, the overhead associated to this operation can be mitigated by combining two techniques:
\begin{itemize}
\item[(i)] an incremental move evaluation approach, using the log-likelihood of the unconstrained subproblem (Lines 9--10) to filter relocation candidates (Line 11), and possibly keeping this solution if it naturally satisfies the assortativity constraints (Lines 12--13);
\item[(ii)] an efficient interior point solver for Problem~(\ref{Convex-DC-SBM01}--\ref{Convex-DC-SBM04}), only used if the relocation candidate was not filtered out due to the previous conditions (Lines 14--19).
\end{itemize}

\begin{algorithm}[htbp]
\caption{Likelihood maximization algorithm}
\begin{algorithmic}[1]
\STATE {\textbf{Input:} Adjacency matrix: $A$, Number of communities: $K$}
\STATE {\textbf{Output:} Network partition: $Z$}
\STATE Initialize a random partition $Z$
\STATE Find $\Omega$ by solving (\ref{Convex-DC-SBM01}--\ref{Convex-DC-SBM04}) for partition $Z$
\STATE Evaluate log-likelihood: $L \leftarrow \log P(A | \Omega, Z)$
\REPEAT
	\FOR{each $i \in \{1,\dots,N\}$ and $r \in \{1,\dots,K\}$}
		\STATE Consider partition $Z^\textsc{R}$ obtained from $Z$ by relocating node $i$ to community $r$
		\STATE Find $\Omega'$ maximizing $\log P(A |\Omega', Z^\textsc{R})$ 
		\STATE Evaluate log-likelihood: $L' \leftarrow \log P(A | \Omega', Z^\textsc{R})$

        \IF{$L' > L$}
		\IF{$\Omega'$ satisfies (\ref{Convex-DC-SBM02}--\ref{Convex-DC-SBM04})}
			\STATE Apply: $\Omega \leftarrow \Omega'$, $Z \leftarrow Z^\textsc{R}$, $L \leftarrow L'$
		\ELSE
			\STATE Find $\Omega''$ by solving (\ref{Convex-DC-SBM01}--\ref{Convex-DC-SBM04}) for partition $Z^\textsc{R}$
			\STATE $L'' \leftarrow \log P(A | \Omega'', Z^\textsc{R})$			
			\IF{$L'' > L$}
				\STATE Apply: \mbox{$\Omega \leftarrow \Omega''$, $Z \leftarrow Z^\textsc{R}$, $L \leftarrow L''$}
			\ENDIF
		\ENDIF
        \ENDIF
	\ENDFOR
\UNTIL{No improving relocation has been identified}
\end{algorithmic}
\label{LocalSearch-SBM}
\end{algorithm}

We use the interior point algorithm of \citet{Domahidi2013} for the solution of each subproblem. When the partition is fixed, the constrained maximization subproblem takes the following form:
\begin{subequations}
\begin{alignat}{3}
\max_{\Omega,\lambda} & \quad \frac{1}{2} \sum_{rs}^{K} \left ( m_{rs} \log(\omega_{rs}) - T_{rs} \omega_{rs}  \right )
 \label{Convex-DC-SBM01}\\
\textrm{s.t.} & \quad \omega_{qq} \geq \lambda \quad \forall q \in \{1,\dots,K\} \label{Convex-DC-SBM02}\\
& \quad \omega_{rs} \leq \lambda \quad \forall r,s \in \{1,\dots,K\}, r \neq s \label{Convex-DC-SBM03}\\
& \quad \omega_{rs} \geq 0 \quad \forall r,s \in \{1,\dots,K\}, \label{Convex-DC-SBM04}
\end{alignat}
\end{subequations}
where $m_{rs}$ represents the number of edges between communities $r$~and~$s$ according to the fixed partition and $T_{rs} =~(\sum_{t}^{K} m_{rt} \sum_{t}^{K} m_{st})/2m$.

\section{Empirical Studies}

We conduct extensive computational experiments on synthetic and real data sets to analyze three aspects of the proposed assortative-constrained DC-SBM (AC-DC-SBM). Firstly, we wish to know under which conditions the assortativity constraints  help to converge to desirable partitions. Secondly, we compare the AC-DC-SBM, the standard DC-SBM and the modularity maximization model in terms of community detection performance. Finally, we apply the AC-DC-SBM to graphs representing brain cortex data, highlighting structures which were not previously detected before and discuss the implications of the different models.

\subsection{Networks Generated From a PPM}

\begin{figure}[htbp]\vspace*{-0.1cm}
\centering
\includegraphics[width=\linewidth]{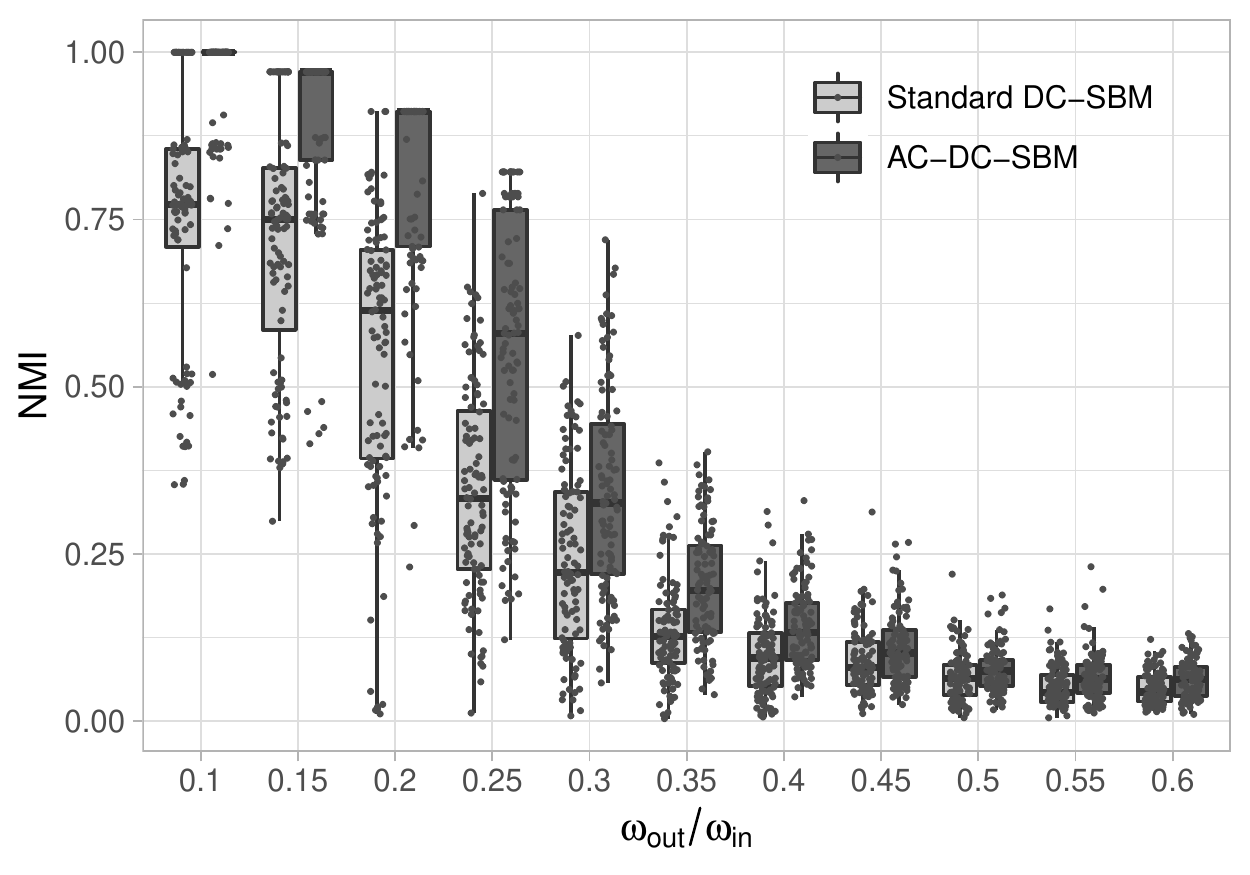}
\caption{Performance of DC-SBM and AC-DC-SBM on networks generated from PPMs with varying degree of assortativity.\label{fig:PhaseTransition}}
\end{figure}

The standard DC-SBM usually finds assortative solutions for assortative networks with a sufficient amount of information. However, it can be trapped into spurious non-assortative local minima on sparse or lightly assortative networks. To limit the number of factors, we conduct this first analysis on data sets generated by a simple PPM (Section~\ref{sec:presentation-PPM}) with $K=4$ blocks, $N = 100$ nodes, an average degree of $16$, and different ratio values for $\omega_\textsc{out}/\omega_\textsc{in}$ representing different assortativity levels. Our goal is to evaluate in which regimes the assortativity constraints are meaningful. Figure~\ref{fig:PhaseTransition} therefore depicts the performance of the standard DC-SBM and of the proposed AC-DC-SBM in terms of normalized mutual information (NMI) \cite{Kvalseth1987}. For each data set and model, we report the results of 100 independent runs from different initial solutions. These results are represented as box plots, where the whiskers extend to 1.5 times the interquartile range.

For the data sets of Figure~\ref{fig:PhaseTransition}, detectability is known to be possible for values of $\omega_\textsc{out}/\omega_\textsc{in}$ smaller than $\approx 0.4$ (see \cite{Decelle2011Phase}). As expected, as the ratio $\omega_\textsc{out}/\omega_\textsc{in}$ increases beyond 0.4, both models are unable to recover the communities. In contrast, when this ratio diminishes below 0.4, the performance of both methods improves, highlighting a phase transition towards a regime where partial recovery is possible. As visible in these experiments, the transition of AC-DC-SBM occurs before that of the standard DC-SBM. For example, when $\omega_\textsc{out}/\omega_\textsc{in} = 0.25$, AC-DC-SBM achieves an average NMI of $0.55$, compared to $0.34$ for DC-SBM. Similarly, when $\omega_\textsc{out}/\omega_\textsc{in} = 0.1$, AC-DC-SBM achieves near-perfect recovery on a much larger proportion of the runs. As such, it appears that the assortativity constraints are useful to guide likelihood maximization algorithms in challenging data sets located within the phase transition regime.

\subsection{Networks Generated From SBMs}

We now repeat the previous experiment on general SBMs, characterized by a larger number of parameters. To compare the results of the DC-SBM and AC-DC-SBM, we generate $50$ synthetic data sets with $N = 100$ nodes and $K = 4$ blocks. For each data set, the $\Omega$ parameters are uniformly sampled in the following intervals:
\begin{align}
\omega_{rr} \in & \enspace \left [ 0.45, 0.55 \right ] & \forall \, r \in \{1,\dots,K\} \label{GenerativeSBM1}\\
\omega_{rs} \in & \enspace \left [ 0, 0.4 \right ] & \forall \, r,s \in \{1,\dots,K\}, r \neq s. \label{GenerativeSBM2}
\end{align}
Each node is allocated to one of the four blocks with equal probability. Then, for each node pair $(i,j)$, a number of edges is generated from a Poisson distribution centered in $\omega_{rs}$, where~$r$ and $s$ represent the blocks of $i$ and $j$.

\begin{figure*}[!htbp]
\centering
\includegraphics[width=\textwidth]{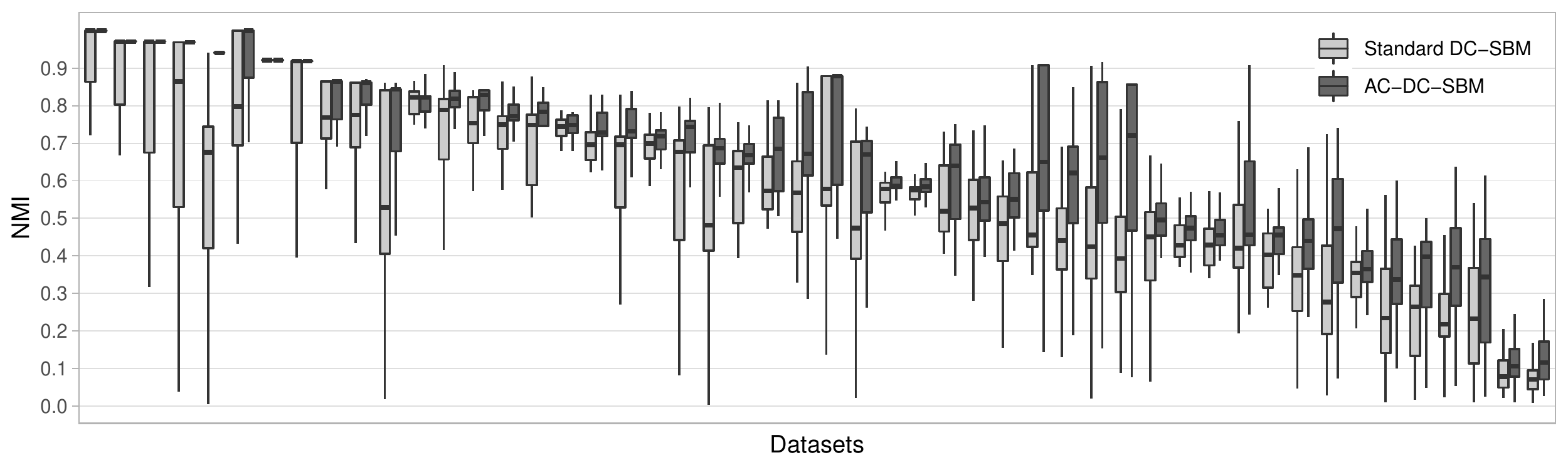}\vspace*{-0.3cm}
\caption{Performance of DC-SBM and AC-DC-SBM on networks generated from general SBMs. The results are ordered by median NMI.\label{fig:Boxplot-05-50}}
\end{figure*}

Figure~\ref{fig:Boxplot-05-50} compares the NMI obtained with the standard DC-SBM and the proposed AC-DC-SBM on these networks. For each network and model, we conduct $50$ independent runs from different initial solutions and report the results as boxplots. AC-DC-SBM obtains on 49 out of 50 datasets a better or equal median NMI than DC-SBM. DC-SBM appears to be very sensible to low-quality local minima. This behavior is particularly visible on the first six data sets presented in the figure. A pairwise Wilcoxon test comparing the average NMI of both methods over the 50 data sets confirms the statistical significance of this difference of performance (with $p=3.9 \times10^{-10}$).

\begin{figure}[htbp]\vspace*{-0.3cm}
    \centering
    \includegraphics[width=\linewidth]{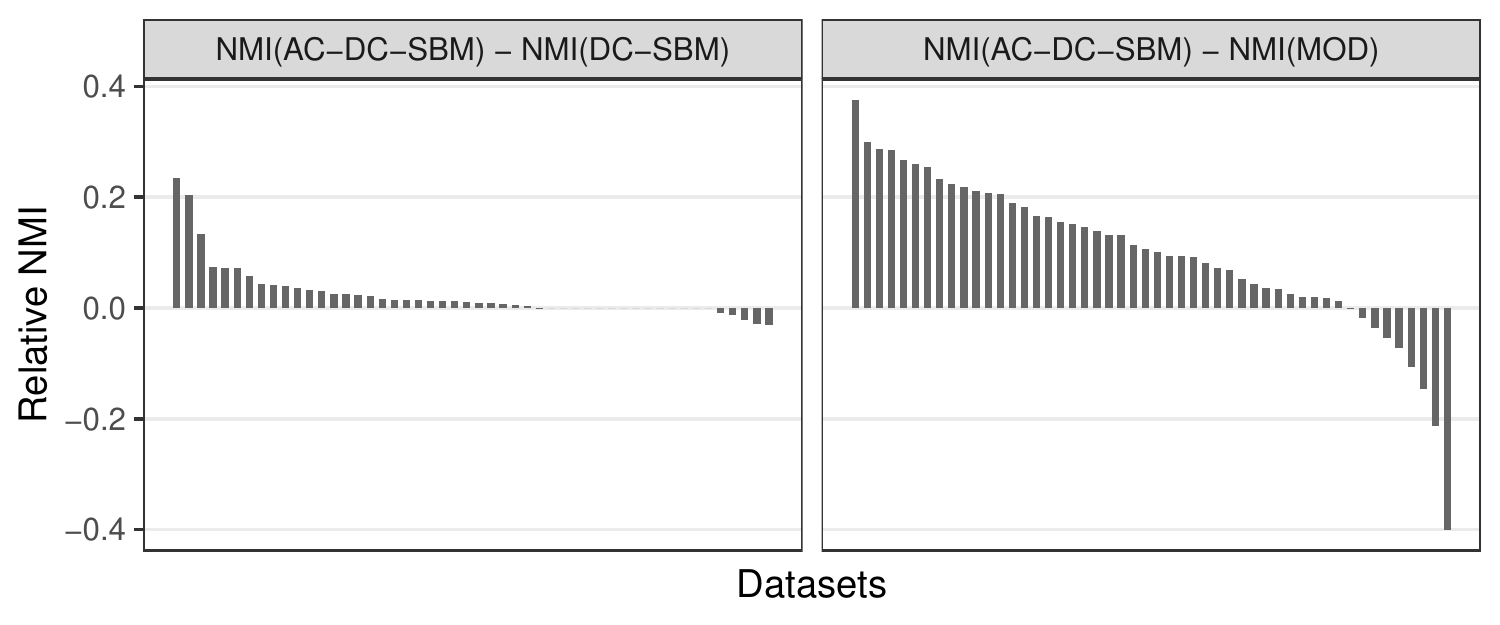}%
    \caption{Relative NMI between the AC-DC-SBM and the standard DC-SBM (left) and modularity-maximization (right). Analysis based on the top 10\% best solutions for each data set.}
\label{fig:Bar-05-5}
\end{figure}

In a second part of this analysis, we filter the set of solutions produced by the methods to focus on the top $10\%$ in terms of likelihood for each data set. This corresponds to a typical use case in which multiple independent runs are performed to avoid local minima. Figure~\ref{fig:Bar-05-5} displays the relative difference between the NMI of the $10\%$ top solutions of the AC-DC-SBM and those of the standard DC-SBM. For the sake of completeness, we repeat the same analysis with the modularity-maximization algorithm. As visible in these results, the best AC-DC-DBM solutions still outperform those of the two other approaches on most data sets. The statistical significance of these observations is confirmed by pairwise Wilcoxon tests (with $p=2.4 \times10^{-5}$ and $p=5.1 \times10^{-6}$ for DC-DBM and modularity maximization, respectively).


\begin{figure}[htbp]
    \centering
	\includegraphics[width=\linewidth]{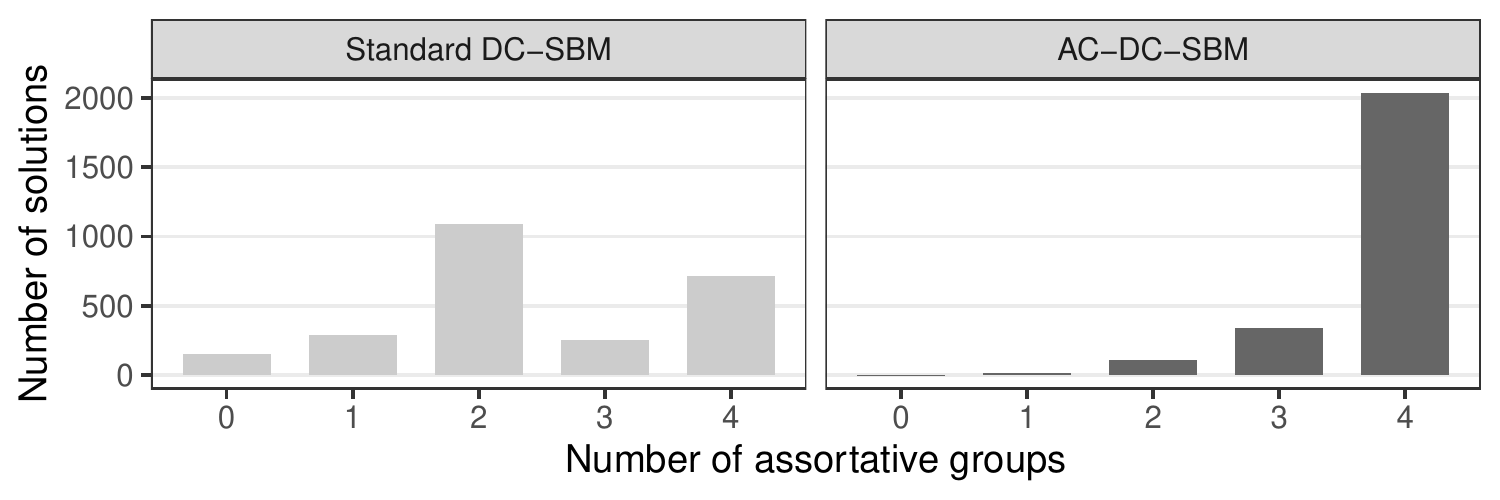}
    \caption{Distribution of the number of assortative communities found by AC-DC-SBM and DC-SBM.}
\label{fig:DensityAss}
\end{figure}

Figure~\ref{fig:DensityAss} finally compares the number of assortative communities found by AC-DC-SBM and DC-SBM. The standard DC-SBM produces much fewer assortative communities in average (2.43 compared to 3.76). As discussed earlier in this article, AC-DC-SBM only enforces constraints on the block-model parameters $\Omega$, and therefore does not systematically guarantee assortative partitions. Yet, non-assortative partitions are heavily penalized from a likelihood perspective and therefore generally avoided. Finally, remark that modularity maximization always produces assortative solutions, but its equivalence to the PPM (with only two parameters driving the distribution of the edges) limits its ability to fit more general SBMs. Among these alternatives, AC-DC-SBM appears to find a trade-off between insufficient and excessive expressiveness.

\subsection{Brain Cortex Networks}

Many real-world networks are known to present assortative structures, e.g., in applications to module or community detection in brain cortex networks, protein-protein interaction, and metabolic networks \cite{Chen2008,Kreimer2008,Ravasz2002, Huss2007}. We analyze in this section the case of the ``cats cortex network'', which is known to have an assortative structure and is divided into four main functional areas: visual, auditory, frontolimbic, and somatosensory-motor duties \cite{Lameu2016}. The network is obtained from the cortico-cortical connectivity pattern described by \cite{Scannell1995}, based on 1139 cortico-cortical connections and 65 cortical areas. As in most community detection tasks, the ground truth in this network is not available. In fact, there is no unique ``correct'' partitioning~\cite{Peel2017}, but different algorithms can allow to highlight different underlying structures.

\begin{figure*}[htbp]\vspace*{-0.3cm}
    \centering
    \subfloat[Standard DC-SBM]{{\includegraphics[width=4.9cm]{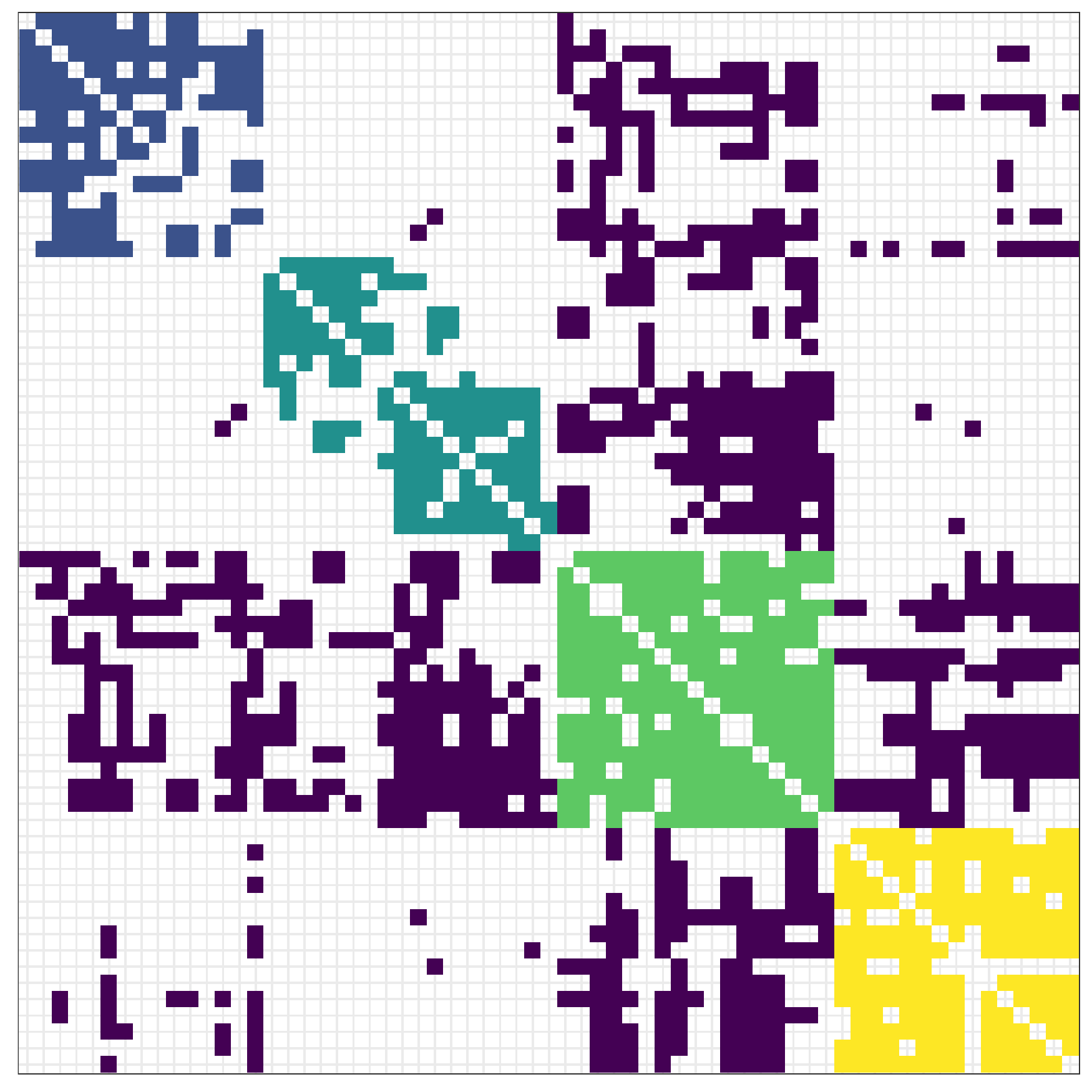} }}%
    \hfill
    \subfloat[AC-DC-SBM]{{\includegraphics[width=4.9cm]{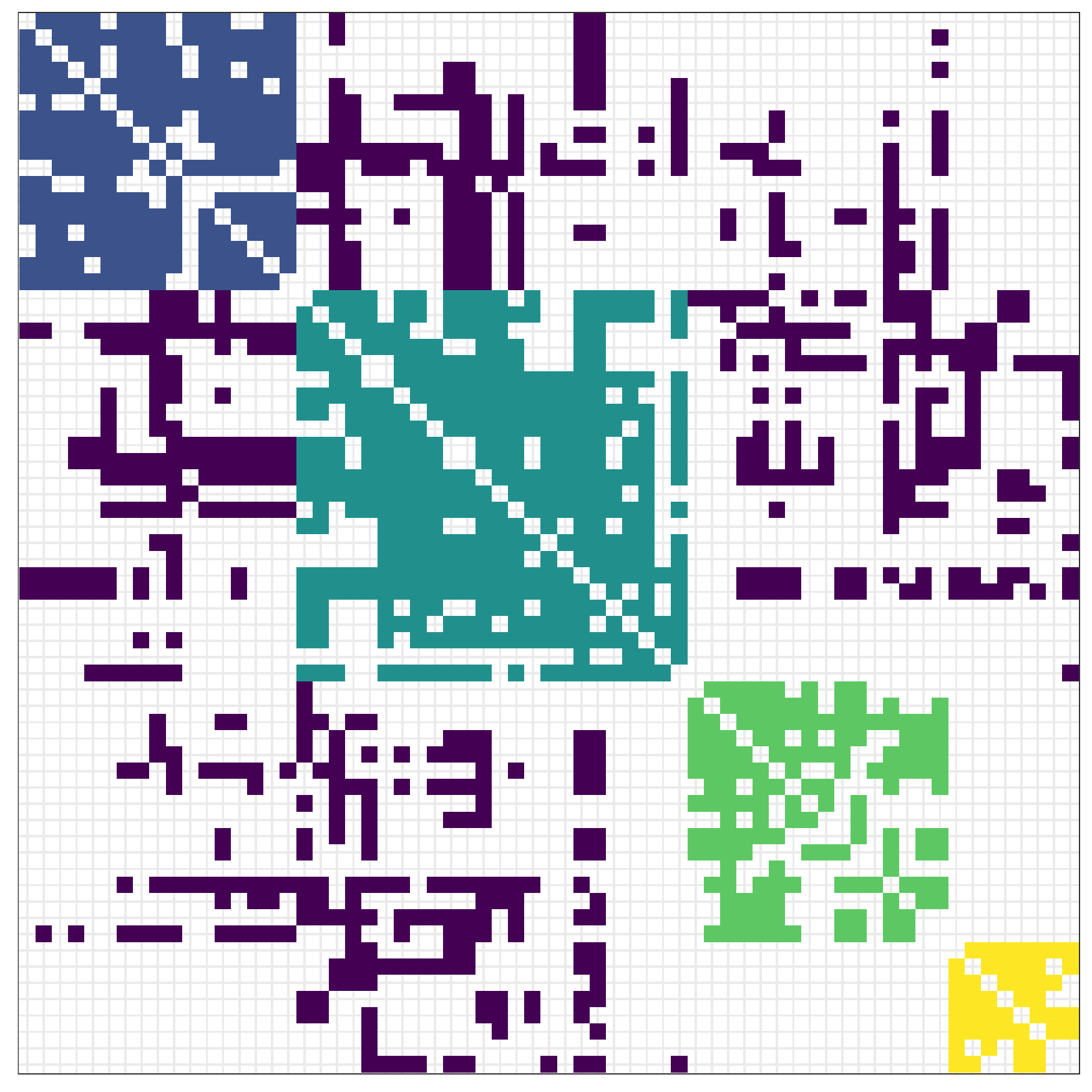} }}%
    \hfill
    \subfloat[Modularity maximization model]{{\includegraphics[width=4.9cm]{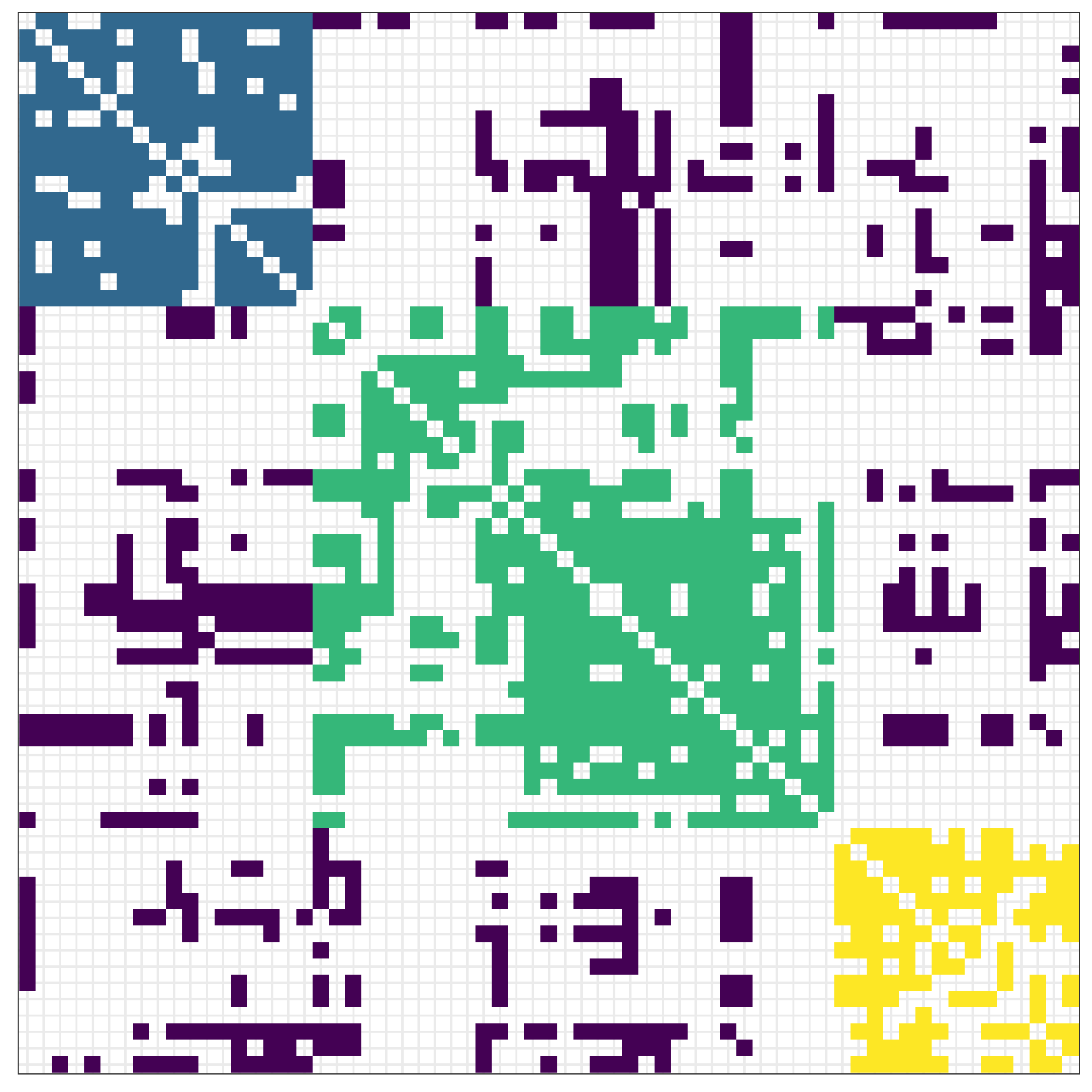} }}%
    \caption{The best among 100 network partitions found by different models in the cats cortex network.}%
    \label{fig:Cats}%
\end{figure*}

Figure~\ref{fig:Cats} reports the communities found with the standard DC-SBM, the AC-DC-SBM and modularity maximization models on this dataset. For each model, we performed 100 optimization runs and registered the best solution (in terms of likelihood or modularity). 

The best solution obtained with the standard DC-SBM is visibly non-assortative. The minimum value found along the $\Omega$ diagonal is 1.5060, whereas the maximum value in the off-diagonal is 1.9050. The size of each group is similar, and one disassortative community acts as a ``hub'' for edges that flow between groups. In contrast, the partition produced by the AC-DC-SBM satisfies the strong assortativity conditions. The minimum value of the $\Omega$ diagonal is 2.0196, and the maximum value in the off-diagonal is 1.7152. This solution includes communities of different sizes with edges which are more evenly distributed between groups. Two mutually-disconnected community pairs are also identified (green-yellow and dark blue-yellow). Finally, the modularity-maximization approach leads to the most assortative partitioning of this network. Yet, since the model does not take $K$ into consideration, this partitioning contains only three groups, contrasting with the four functional areas which were originally expected.

\section{Conclusions}
\label{sec:Conclusions}

Assortativity constraints arise as a natural approach to guide maximum-likelihood algorithms away from spurious local minima on networks which have a presupposed assortative structure. In this work, we have shown that these constraints can be effectively handled with tailored local optimization and interior point methods. Our experiments show that the resulting AC-DC-SBM significantly outperforms unconstrained community detection methods in lightly assortative graphs, especially in regimes which are close to the detectability threshold. In these circumstances, the classic SBM has a strong tendency to converge towards non-assortative solutions, while the modularity maximization model does not generalize well to graphs in which the number of edges between groups widely varies. On the practical example of a brain cortex network, the proposed AC-DC-SBM reveals drastically different community structures which were not identified by other algorithms.

The research perspectives related to this work are numerous. We recommend to further evaluate the impact of assortativity constraints on known phase transitions and thresholds. We also recommend to investigate different algorithmic paradigms to improve the solution of this constrained maximum likelihood formulation, and to pursue the study of the AC-DC-SBM in a wider range of application contexts.

\end{document}